\documentclass[%
 reprint,
 amsmath,amssymb,
 aps,
]{revtex4-1}

\usepackage{graphicx}
\usepackage{dcolumn}
\usepackage{bm}

\usepackage{framed}

\begin{document}

\preprint{APS/123-QED}

\title{Hidden symmetries in $N$-layer dielectric stacks}

\author{Haihao Liu$^{1,2}$} 
\author{M. Shoufie Ukhtary$^{1}$}
\email{shoufie@flex.phys.tohoku.ac.jp}
\author{Riichiro Saito$^{1}$}
\affiliation{$^{1}$Department of Physics, Tohoku University, Sendai 980-8578, Japan.\\$^{2}$Department of Materials Science and NanoEngineering, Rice University, Houston, TX 77005-1892, USA.}

\date{\today}

\begin{abstract}
The optical properties of a multilayer system of dielectric media with arbitrary $N$ layers is investigated. Each layer is one of two dielectric media, with thickness one-quarter the wavelength of light in that medium, corresponding to a central frequency. Using the transfer matrix method, the transmittance $T$ is calculated for all possible $2^N$ sequences for small $N$. Unexpectedly, it is found that instead of $2^N$ different values of $T$ at the central frequency ($T_0$), there are either $(N/2+1)$ or $(N+1)$ discrete values of $T_0$ for even or odd $N$, respectively. We explain the high degeneracy in the $T_0$ values by defining new symmetry operations that do not change $T_0$. Analytical formulae were derived for the $T_0$ values and their degeneracy as functions of $N$ and an integer parameter for each sequence we call ``charge''. Additionally, the bandwidth of the transmission spectra at $f_0$ is investigated, revealing some asymptotic behavior at large $N$.
\begin{description}
\item[PACS numbers]
\verb+42.25.Bs,78.20.Bh,78.67.Pt+.
\end{description}
\end{abstract}

\maketitle


\section{Introduction}

The advances in electronics have enabled us to control electron transport through materials, allowing us to develop electronic devices, such as transistors and diodes. However, due to the temporal and spatial limitations of electrons, transporting information using electrons for long distance is not efficient. Light can be used for such purposes as the carrier of information instead of electrons~\cite{saleh1991,joannopoulos2011}. Nowadays, controlling the propagation of light has been the main subject in optical engineering, which is known as photonics~\cite{saleh1991,joannopoulos2011,joannopoulos1997photonic,krauss1999photonic}. Similar to electronics, in photonics, one tries to modify how light propagates through materials, including how one can allow or prevent the propagation of light, or localize the light~\cite{joannopoulos2011,chigrin1999observation,orfanidis2002electromagnetic}, which can be useful to amplify the electric field. To achieve this, one can use a multilayer system consisting of $N$ layers of dielectric media varying in one-dimension. We will refer to this system as a multilayer stack. By manipulating the sequence of dielectric media in one dimension, one can control how light propagates through it~\cite{saleh1991,joannopoulos2011,lusk2001omnidirectional,orfanidis2002electromagnetic}.

Previous studies have shown that sequences of dielectric media with a periodic structure, known as photonic crystals (PCs), or generated based on fractal patterns can control the propagation properties of light through the multilayer stack, such as transmission (T) spectra, group velocity and dispersion~\cite{xu2010one,tavakoli2014one,endo2011tunneling,hattori1994photonic,krauss1999photonic}. One can expect some desirable optical properties, such as a very sharp and localized peak in the T spectrum or high electric field enhancement at a specific point, which allow us to develop optical filters, widely known as Fabry-Perot resonators~\cite{orfanidis2002electromagnetic,van1985multimirror,banning1947practical}, and optical switches~\cite{tavakoli2014one}. One can also realize perfect dielectric mirrors based on a multilayer stack, which are known as Bragg reflectors~\cite{orfanidis2002electromagnetic,fink1998dielectric,winn1998omnidirectional,turner1966multilayer}, as well as structures based on them called Bragg-grating filters~\cite{wei1997phase,erdogan1997fiber,bakhti1997design,zengerle1995phase}. Furthermore, if one adds a conducting layer inside the multilayer stack, such as metal or graphene, one can also control the absorption of the light intensity as a function of Fermi energy of the metal layer~\cite{ukhtary2015,reynolds2016,bonaccorso2010graphene,harada2016giant} and we can expect enhancement of the absorption due to the high electric field enhancement inside the multilayer system. However, these previous studies focus only on some specific sequences of dielectric media, such as alternating and periodic sequences or Fibonacci and Cantor sequences~\cite{xu2010one,tavakoli2014one,endo2011tunneling,hattori1994photonic,monsoriu2005cantor}. The general optical properties for \textit{any arbitrary} sequence in a multilayer stack have not yet been discussed as far as we know, simply because we did not have a systematic analysis to understand the phenomena for any arbitrary sequence.

In this study, we investigate the optical properties of a multilayer system consisting of arbitrary sequences of $N$ layers, in particular the transmittance of light $T$ through the system. In this work, the $N$ layers are made of two kinds of dielectric media. In contrast to the previous studies, which discussed only very specific sequences, we calculate $T$ for all possible $2^N$ sequences. Hence, our system includes all previously mentioned sequences. One might think that there is no pattern in $T$ for an arbitrary sequence of the $N$-layer stack. In this work we found that instead of $2^N$ different values, at a particular central (or resonant) frequency $f_0$, there are either $(N/2+1)$ or $(N+1)$ discrete values for even or odd number $N$, respectively, provided we select the thickness of each layer to be one-quarter the wavelength of light in that layer corresponding to $f_0$. This high degeneracy generally implies the existence of hidden symmetry operations for exchanging the dielectric layers in the stack. In particular, we will define a new integer parameter called ``charge'' which is invariant for the operations. We will show that all $T$ values at $f_0$ are given by the ``charge'' and understanding the origin of these hidden symmetries and patterns can be useful for finding and designing optimal sequences, especially for systems with large $N$. 

Our paper is organized as follows. In Sec. II we will describe our method to calculate the $T$ of a multilayer system with arbitrary $N$ layers of dielectric media. In Sec. III we will show our results and explain the symmetry operations of the multilayer system. We will also provide the analytical formula of $T$ at $f_0$ as a function of ``charge'' in Sec. III, as well as briefly discuss the bandwidth of the $T$ spectra (i.e. how sharp the peak at $f_0$ is). We will give our conclusion in Sec. IV. All the mathematical proofs are given in the Appendix.

\section{Method}

In Fig.~\ref{fig1}, we show a schematic picture of our multilayer system consisting of $N$ layers of dielectric media where the $i$-th layer, $L_i$, is one of two dielectric media that are labeled by A and B, with refraction indices $n_{\text{A}}$ and $n_{\text{B}}$, respectively. The thickness of $L_i$ is selected as $\ell_i=\lambda_0/4n_i$, where $\lambda_0=c/f_0$ is the wavelength of light in vacuum with frequency $f_0$, which is chosen as the central frequency, and $n_i$ is either $n_{\text{A}}$ and $n_{\text{B}}$. Hence, we have $2^N$ possible sequences of $L_1L_2...L_N$, for example with $N=6$, we have 64 different possible sequences.

\begin{figure}[h]
\centering\includegraphics[width=7cm]{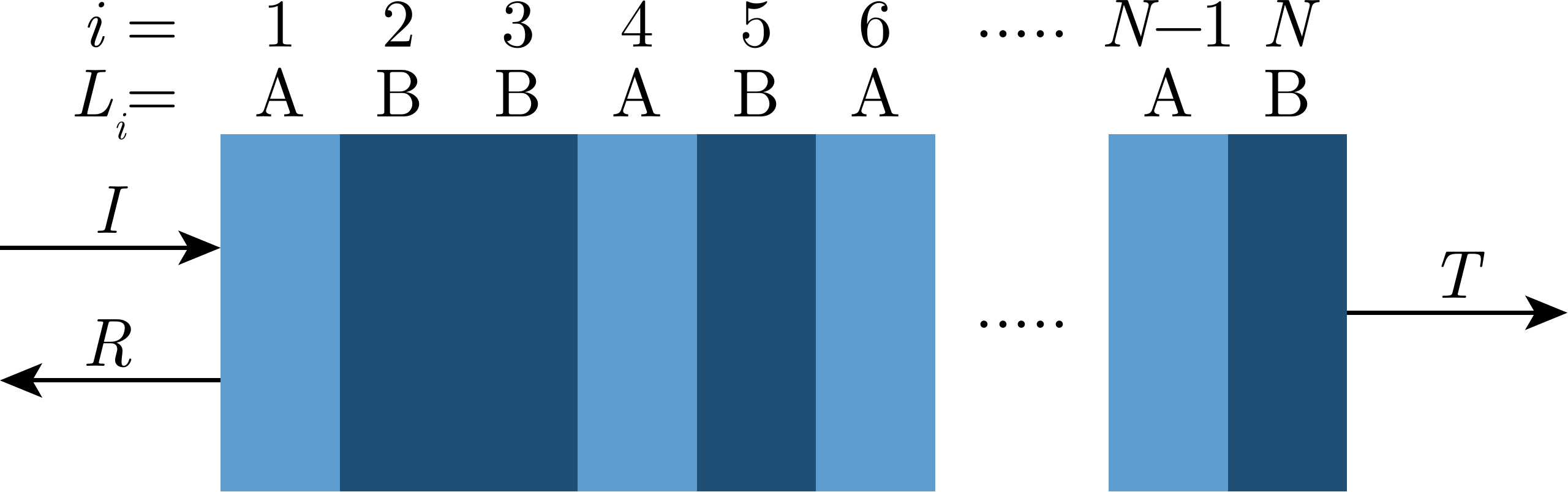}
\caption{Schematic picture of multilayer system. A multilayer system consists of $N$ layers of dielectric media where the $i$-th layer, $L_i$, is one of two dielectric media, A and B, shown as two different colors, therefore there are $2^N$ possible sequences. Reflectance and transmittance of light are shown as $R$ and $T$ respectively.}
\label{fig1}
\end{figure}

We assume that the incident light $I$ is normal to the surface of the layer. The reflectance and transmittance of light, $R$ and $T$ respectively, can be calculated by the transfer matrix method~\cite{reynolds2016,bendickson1996analytic}. By using the transfer matrix method, we can relate the electromagnetic (EM) fields of light between any two different positions without knowing the multiple reflection processes between them in detail. This method has been used in previous studies of propagation of a wave inside varying media~\cite{reynolds2016,xu2010one,tavakoli2014one,endo2011tunneling,jonsson1990solving,ko1988matrix}. We will briefly show the transfer matrix method as below.

\begin{figure}[h]
\centering\includegraphics[width=8cm]{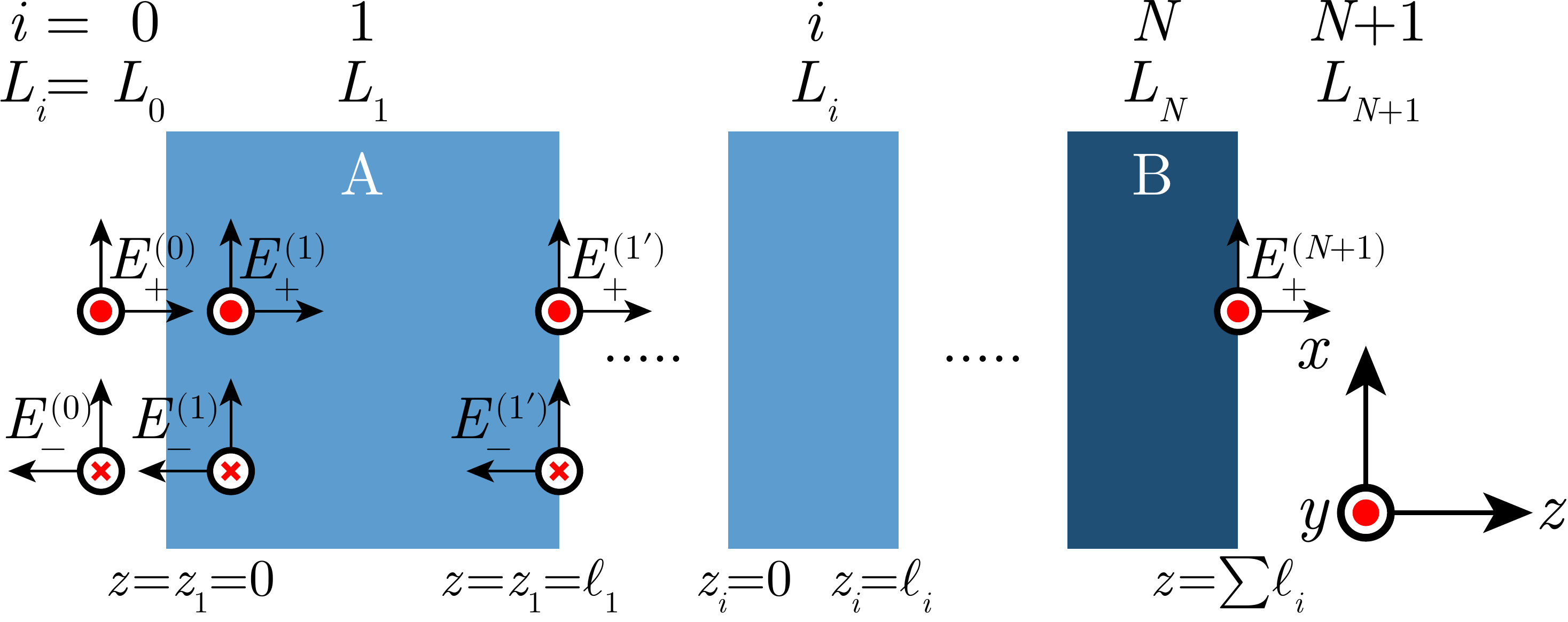}
\caption{Schematic of light propagation from vacuum (0) through $N$-layer dielectric stack. $E^{(0)}_{\pm}$ and $E^{(1)}_{\pm}$ (in general $E^{(i')}_{\pm}$ and $E^{(i+1)}_{\pm}$) should both be exactly on the boundary, which cannot be easily represented.}
\label{fig2}
\end{figure}

In Fig.~\ref{fig2} we define the electric field of left- $(-)$ and right- $(+)$ going waves from $z=0$ to $z=\sum_{i=1}^{N}\ell_i$. The light propagates in the $z$-direction and the electric field is chosen to be in the $x$-direction. In this case, the magnetic field lies in the positive (negative) $y$-direction which we show as red dots (crosses) in Fig.~\ref{fig2}. $E^{(i)}_{+}$ and $E^{(i)}_{-}$ are electric field amplitude at the leftmost edge of $L_i$ for right- and left-going waves respectively. In this paper, $L_0$ and $L_{N+1}$ are taken as vacuum. Therefore, $E^{(0)}_{+}$ and $E^{(0)}_{-}$ denote the incident and reflected electric fields, respectively, while $E^{(N+1)}_{+}$ is the transmitted field.  

The electric field in $L_i$ as a function of $z_i$ (local $z$-coordinate, $z_i$=0 at the leftmost edge of $L_i$) is given by
\begin{align}
E^{(i)}(z_i) &= E_+^{(i)}e^{ik_iz_i}+E_-^{(i)}e^{-ik_iz_i},
\label{eq:ef}
\end{align}
where $k_i=2\pi n_i/\lambda$ is the wavevector of light with wavelength $\lambda$ in $L_i$. Eq.~(\ref{eq:ef}) means that the electric field can be written as a superposition of right- and left-going electromagnetic waves. The magnetic field is related to the electric field by the following equation: $H^{(i)}(z_i)=i\omega \varepsilon_0 n_i^2 \int{E^{(i)} dz_i}$. Thus the magnetic field in $L_i$ as a function of $z_i$ is given by
\begin{align}
H^{(i)}(z_i)&=\frac{\omega\varepsilon_0 n_i^2}{k_{i}}\left(E_{+}^{(i)}e^{ik_iz_i}-E_{-}^{(i)}e^{-ik_iz_i}\right).
\label{eq:mf}
\end{align}

The total electric and magnetic fields are continuous at the interface between $L_i$ and $L_{i+1}$, so in terms of amplitude, $E^{(i)} = E^{(i+1)}$ and $H^{(i)} = H^{(i+1)}$. Let us take for example the interface between layer 0 and layer 1 shown in Fig.~\ref{fig2}. Using Eqs.~(\ref{eq:ef})-(\ref{eq:mf}) and the above, we get
\begin{align}
E_+^{(0)}+E_-^{(0)}&=E_+^{(1)}+E_-^{(1)},
\label{eq:bound1}\\
E_+^{(0)}-E_-^{(0)}&=\frac{k_0}{k_1}\left(\frac{n_1}{n_2}\right)^2\left(E_+^{(1)}-E_-^{(1)}\right).
\label{eq:bound2}
\end{align}
From Eq.~(\ref{eq:bound1}) and ~(\ref{eq:bound2}), we can form a matrix that relates the electric fields across the interface,
\begin{align}
\left[ \begin{array}{c}
E_{+}^{(0)}\\
E_{-}^{(0)}
\end{array} \right]=
\frac{1}{2}\left[ \begin{array}{cc}
1+\beta_0 & 1-\beta_0\\
1-\beta_0 & 1+\beta_0
\end{array} \right]
\left[ \begin{array}{c}
E_{+}^{(1)}\\
E_{-}^{(1)}
\end{array} \right],
\label{eq:m}
\end{align} 
where $\beta_i$ denotes
\begin{align}
\beta_i=\frac{k_{i}}{k_{i+1}}\left(\frac{n_{i+1}}{n_i}\right)^2=\frac{n_{i+1}}{n_i}.
\label{eq:beta}
\end{align}

After entering $L_1$, the right-going light propagates through $L_1$ until it hits another interface with $L_2$. During the propagation inside $L_1$ (from $z=0$ to $z=\ell_1$), the electric field changes only by its phase. From Eq.~(\ref{eq:ef}), we can form another matrix that relate the electric field at $z=0$ and $z=\ell_1$ inside $L_1$,
\begin{align}
\left[ \begin{array}{c}
E_{+}^{(1)}\\
E_{-}^{(1)}
\end{array} \right]=
\left[ \begin{array}{cc}
e^{-ik_{1}\ell_1} & 0\\
0 & e^{ik_{1}\ell_1}
\end{array} \right]
\left[ \begin{array}{c}
E_{+}^{(1')}\\
E_{-}^{(1')}
\end{array} \right] ,
\label{eq:p}
\end{align}
where $E_{+}^{(1')}$ and $E_{-}^{(1')}$ are the electric fields in $L_1$ at $z=\ell_1$. We can combine the matrices of Eq.~(\ref{eq:m}) and Eq.~(\ref{eq:p}) to get
\begin{align}
\left[ \begin{array}{c}
E_{+}^{(0)}\\
E_{-}^{(0)}
\end{array} \right]=
M_0 P_1
\left[ \begin{array}{c}
E_{+}^{(1')}\\
E_{-}^{(1')}
\end{array} \right],
\label{eq:tf1}
\end{align}
where
\begin{align}
M_i & = \frac{1}{2}\left[ \begin{array}{cc} 1+\beta_i & 1-\beta_i~\\
1-\beta_i & 1+\beta_i
\end{array} \right], \label{eq:mi} \\
P_i & = \left[ \begin{array}{cc}
e^{-ik_{i}\ell_{i}} & 0\\
0 & e^{ik_{i}\ell_{i}}
\end{array} \right].
\end{align}
$M_i$ and $P_i$ are called matching and propagation matrices, respectively. The product of $M_0P_1$ in Eq.~(\ref{eq:tf1}) is known as the transfer matrix~\cite{reynolds2016}.

This transfer matrix describes the propagation of incident light from vacuum $L_0$ through $L_1$. If we have multiple layers, we can continue the multiplication of $M_{i-1}$ and $P_i$ for $i=2,\dots,N$. We can write the transfer matrix for an $N$-layer system as follows,
\begin{align}
\left[ \begin{array}{c}
E_{+}^{(0)}\\
E_{-}^{(0)}
\end{array} \right]=
M_0P_1M_1\cdots P_NM_N
\left[ \begin{array}{c}
E_{+}^{(N+1)}\\
0
\end{array} \right],
\label{eq:tf2}
\end{align}
where we do not expect any left-going light coming to the system at $L_{N+1}$. The product of $M_{i-1}$ and $P_i$ in Eq.~(\ref{eq:tf2}) can be expressed by a $2\times 2$ matrix as follows,
\begin{align}
\left[ \begin{array}{c}
E_{+}^{(0)}\\
E_{-}^{(0)}
\end{array} \right]&=
\left[ \begin{array}{cc}
a & b\\
c & d
\end{array} \right]
\left[ \begin{array}{c}
E_{+}^{(N+1)}\\
0
\end{array} \right]
\end{align}
which gives us the transmittance of light $T$,
\begin{align}
T&=\left|\frac{E_{+}^{(N+1)}}{E_{+}^{(0)}}\right|^2\nonumber\\
&=\left|\frac{1}{a}\right|^2.
\label{eq:t}
\end{align}
Using the transfer matrix method, we can calculate the $T$ for any arbitrary sequence. In the next section we show numerically calculated $T$'s for all $2^N$ different sequences. 

\section{Results and discussion}

\subsection{Transmittance of light}
The transmittance $T$ as a function of incident frequency $f$ was calculated by MATLAB. We choose dielectric constants (or relative permittivities) $\varepsilon_{\text{A}}$ and $\varepsilon_{\text{B}}$ of the two dielectric media to be 4 and 2.25 respectively. Index of refraction is $n = \sqrt{\mu\varepsilon}$, with relative permeability $\mu \approx 1$, so $n_{\text{A}}$ and $n_{\text{B}}$ are taken to be $\sqrt{\varepsilon_{\text{A}}}=2$ and $\sqrt{\varepsilon_{\text{B}}}=1.5$, respectively, and used throughout this paper for simplicity. Examples of common real materials with refractive indices very close to these include silicon nitride ($\mathrm{Si_3N_4}$) for $n_{\text{A}}$~\cite{luke2015broadband} and silica or acrylic glass for $n_{\text{B}}$. As mentioned before, the thickness of each layer is one-quarter the central wavelength $\lambda_0$ in that layer, i.e. $\ell_i=\lambda_0/4n_i$. If we choose $f_0=2$ THz, then $\ell_{\text{A}}=18.75~\mathrm{\mu}$m and $\ell_{\text{B}}=25~\mathrm{\mu}$m. Note as a matter of convention that in this paper, the sequence $L_i$'s are represented by a string of A's and B's, such as ABAABBA.

\begin{figure}[h]
\centering\includegraphics[width=8cm]{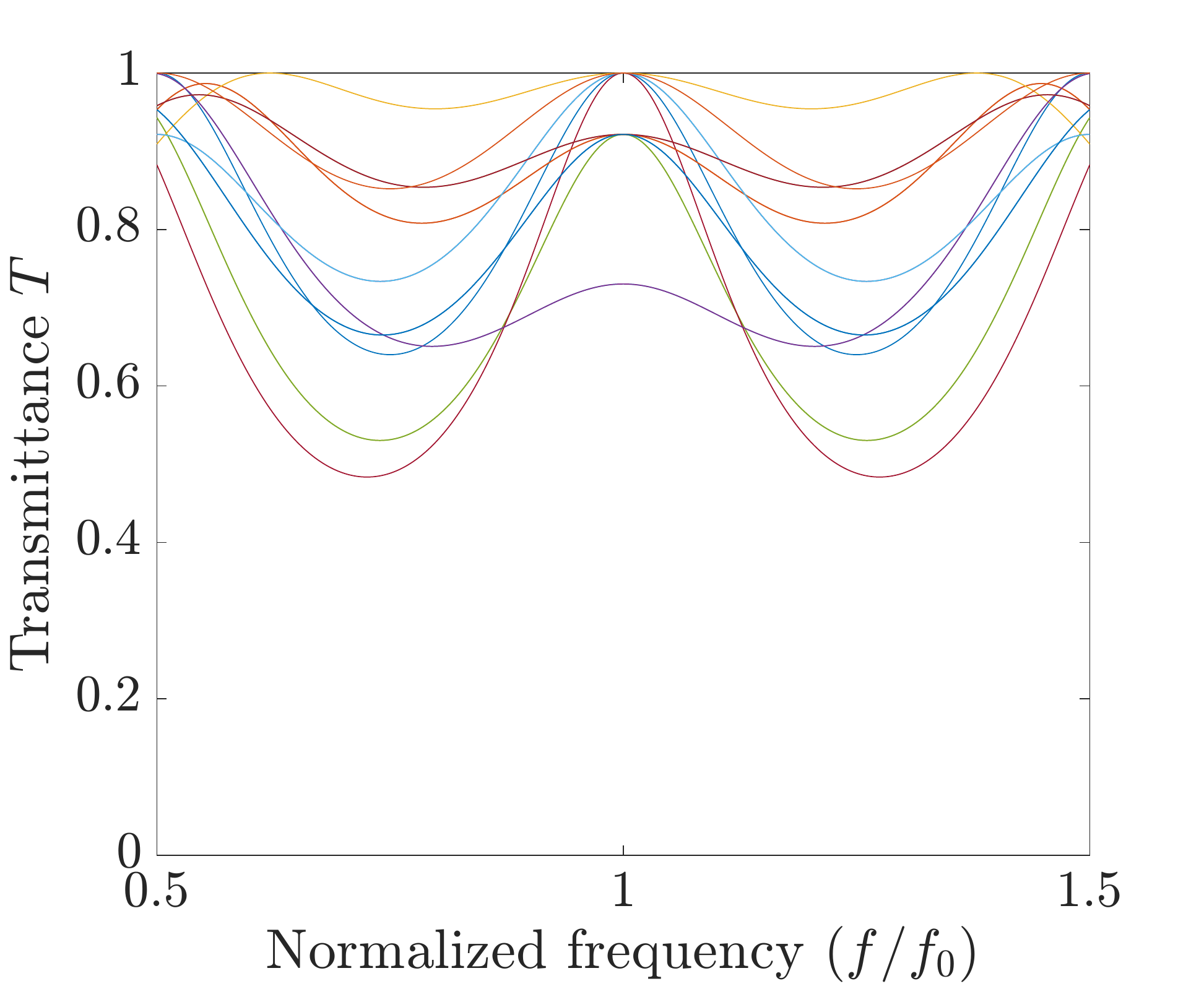}
\caption{For $N = 4$, $T$ spectra for all $2^4 = 16$ possible sequences. Due to the mirror symmetry, there are only 10 unique spectra.}
\label{fig:tspec4}
\end{figure}

\begin{figure}[h]
\centering\includegraphics[width=7.95cm]{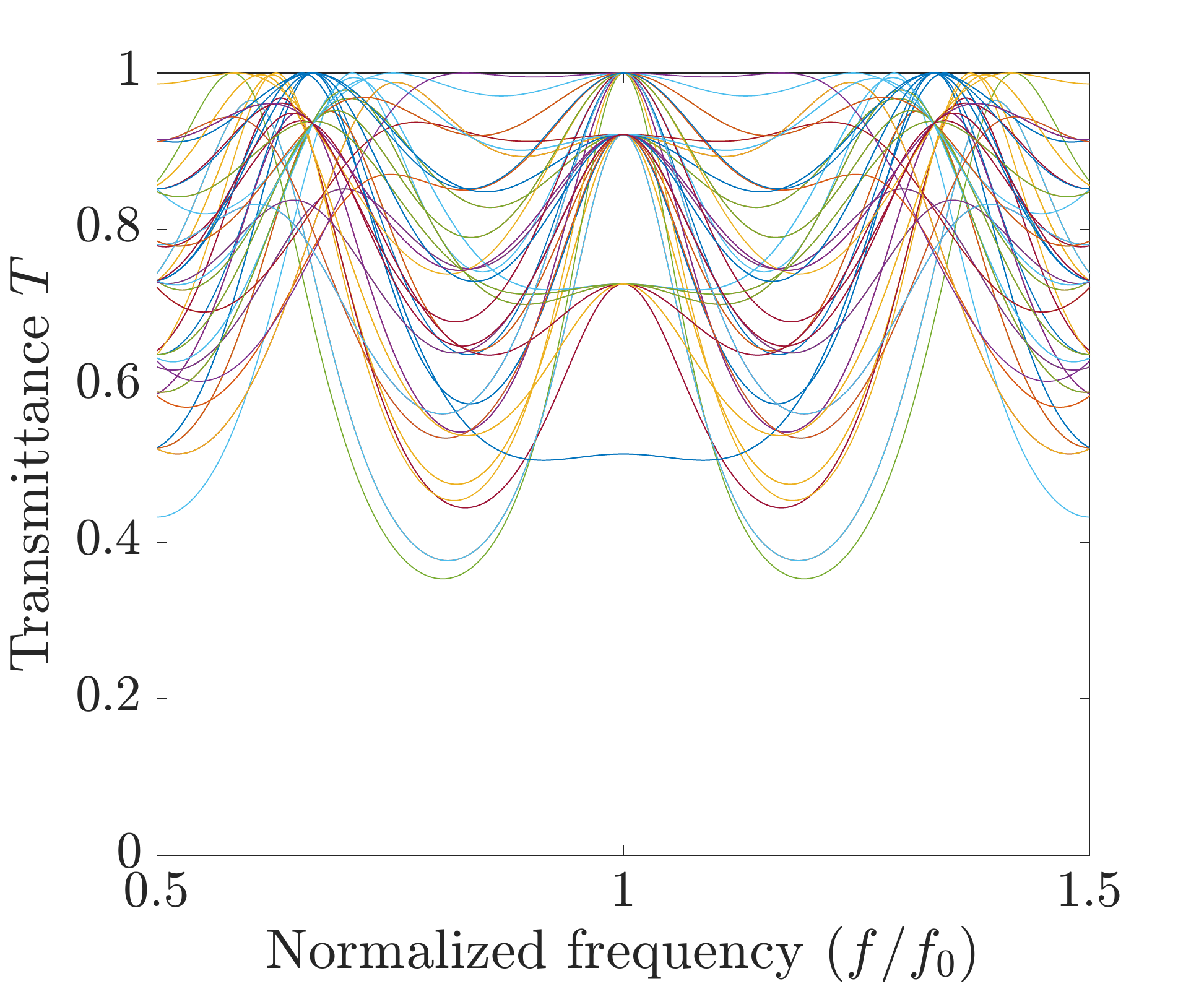}
\caption{For $N = 6$, $T$ spectra for all $2^6 = 64$ possible sequences. Due to the mirror symmetry, there are only 36 unique spectra.}
\label{fig:tspec}
\end{figure}

In Fig.~\ref{fig:tspec4}, $T$ is plotted as a function of frequency normalized to central frequency $f_0$ for all 16 possible 4-layer sequences. We also show the same plot for $6$-layer sequences in Fig.~\ref{fig:tspec}. It is noted that the shape of the spectra does not change for different $f_0$'s, since  $\ell_i$'s also change accordingly, hence why we plot $T$ as a function of $f/f_0$.

The first thing that is noticed in Fig.~\ref{fig:tspec4} is that there are not actually 16 unique spectra, but only 10, by counting the number of curves on the graph. Upon investigation, it is realized that sequences that are mirrored versions of each other, e.g. AABA and ABAA, would produce identical $T$ spectra. This is not too surprising, since light propagating through the sequence one way is essentially equivalent to light propagating through the mirrored sequence the other way (or the time-reversal symmetry of $T$~\cite{matuschek1997exact}). Detailed proofs of this mirror symmetry can be found in Appendix B1.

It is also noticed that the curves seem to converge at three points at $f_0$. To investigate this further, the $T$ spectra for all 64 possible 6-layer sequences are calculated, and can be seen in Fig.~\ref{fig:tspec}. We found that due to mirror symmetry, there are only 36 unique spectra. By considering the number of symmetric or ``palindromic" sequences, which are invariant under mirror symmetry, we determine the number of unique spectra for an $N$-layer system to be:
\begin{align}
&\textrm{Even}~N\nonumber\\
2^{\frac{N}{2}}+&\frac{1}{2}\left(2^{N}-2^{\frac{N}{2}}\right)\\
&\textrm{Odd}~N\nonumber\\
2^{\frac{N+1}{2}}+&\frac{1}{2}\left(2^{N}-2^{\frac{N+1}{2}}\right).
\label{eq:num}
\end{align}

With the greater number of spectra, it is clear that they are converging to 4 $T$'s at $f_0$, in the case of $N = 6$. This cannot be accounted for only by ``mirror symmetry'', because different spectra give the same $T$ at $f_0$, hereafter denoted $T_0$. This is a surprising result, because by changing one layer from A to B, for example, one would expect the complex interactions of internal multiple reflections to completely change, and thus have a completely different $T$. Indeed, this is the behavior at frequencies other than $f_0$, where we see many non-degenerate spectra. The high degeneracy at $f_0$ implies there are hidden symmetries besides simple mirror symmetry to be found  in the sequences, giving rise to the degeneracy.

In order to begin finding patterns and understanding this phenomenon, the number of unique $T_0$ values as a function of $N$ is calculated and listed for $N=1$ through 12, and shown in Table~\ref{tab:uniqt}.

\begin{table}[h]
\centering
\begin{tabular}{ccccccccccccc}
\hline\hline
$N$ & 1 & 2 & 3 & 4 & 5 & 6 & 7 & 8 & 9 & 10 & 11 & 12 \\
\hline
No. of $T_0$ values & 2 & 2 & 4 & 3 & 6 & 4 & 8 & 5 & 10 & 6 & 12 & 7 \\
$N/2+1$~(Even $N$) & & 2 & & 3 & & 4 & & 5 & & 6 & & 7 \\
$N+1$~(Odd $N$)& 2 & & 4 & & 6 & & 8 & & 10 & & 12 & \\
\hline\hline
\end{tabular}
\caption{Number of unique $T_0$ values for $N=1$ to 12.}
\label{tab:uniqt}
\end{table}

There is a clear pattern in the number of $T_0$ values, but different patterns for even and odd $N$. It was conjectured that the number of $T_0$ values for even $N$ is $N/2+1$, and for odd $N$, $N+1$, which are also shown in Table~\ref{tab:uniqt}. These numbers are proved in subsection C Formula for $T_0$.

The sequences which all give the same $T_0$ are manually tabulated in Table~\ref{tab:6seq} for $N = 6$, all 64 sequences. The number of sequences at each $T_0$, which we may call the degeneracy, is listed as well. As an example, there are 20 6-layer sequences that gave a $T_0$ of 1.0 (perfect transmittance). These are listed in the first part of Table~\ref{tab:6seq}.

It is not at all obvious what these sequences with the same $T_0$ have in common, i.e. how they are related by symmetry operations like mirroring, hence hidden symmetries is an apt name. As we begin to find these symmetries, it is clear that even and odd $N$ do not have the same symmetries. Since the symmetries for even $N$ seemed less elusive, we focus our efforts on finding all the hidden symmetries for even $N$ in the next subsection that can explain how sequences give the same $T_0$. These 20 sequences also serve as a prototypical example demonstrating why all the symmetry operations are needed.

\begin{table}[h]
\begin{tabular}{ccccc}
\hline\hline
\multicolumn{2}{l}{$T_0$ = 1.0} & \multicolumn{2}{l}{degeneracy = 20} & \multicolumn{1}{l}{$|q|=0$}  \\
AAAAAA & AAAABB & AAABBA & AABAAB & AABBAA \\
AABBBB & ABAABA & ABBAAA & ABBABB & ABBBBA \\
BAAAAB & BAABAA & BAABBB & BABBAB & BBAAAA \\
BBAABB & BBABBA & BBBAAB & BBBBAA & BBBBBB \\
\hline
\multicolumn{2}{l}{$T_0$ = 0.9557} & \multicolumn{2}{l}{degeneracy = 30} & \multicolumn{1}{l}{$|q|=1$} \\
AAAAAB & AAAABA & AAABAA & AAABBB & AABAAA \\
AABABB & AABBAB & AABBBA & ABAAAA & ABAABB \\
ABABBA & ABBAAB & ABBABA & ABBBAA & ABBBBB \\
BAAAAA & BAAABB & BAABAB & BAABBA & BABAAB \\
BABBAA & BABBBB & BBAAAB & BBAABA & BBABAA \\
BBABBB & BBBAAA & BBBABB & BBBBAB & BBBBBA \\
\hline
\multicolumn{2}{l}{$T_0$ = 0.8374} & \multicolumn{2}{l}{degeneracy = 12} & \multicolumn{1}{l}{$|q|=2$} \\
AAABAB & AABABA & ABAAAB & ABABAA \\
ABABBB & ABBBAB & BAAABA & BABAAA \\
BABABB & BABBBA & BBABAB & BBBABA \\
\hline
\multicolumn{2}{l}{$T_0$ = 0.68} & \multicolumn{2}{l}{degeneracy = 2} & \multicolumn{1}{l}{$|q|=3$} \\
ABABAB & BABABA \\
\hline\hline
\end{tabular}
\caption{Sequences that give the same $T_0$ values for $N= 6$. $|q|$ is the total charge of the sequence (see subsection C). The degeneracy is at $f_0$. }
\label{tab:6seq}
\end{table}

\subsection{Symmetry for even $N$}
\begin{figure}[t]
\centering\includegraphics[width=6cm]{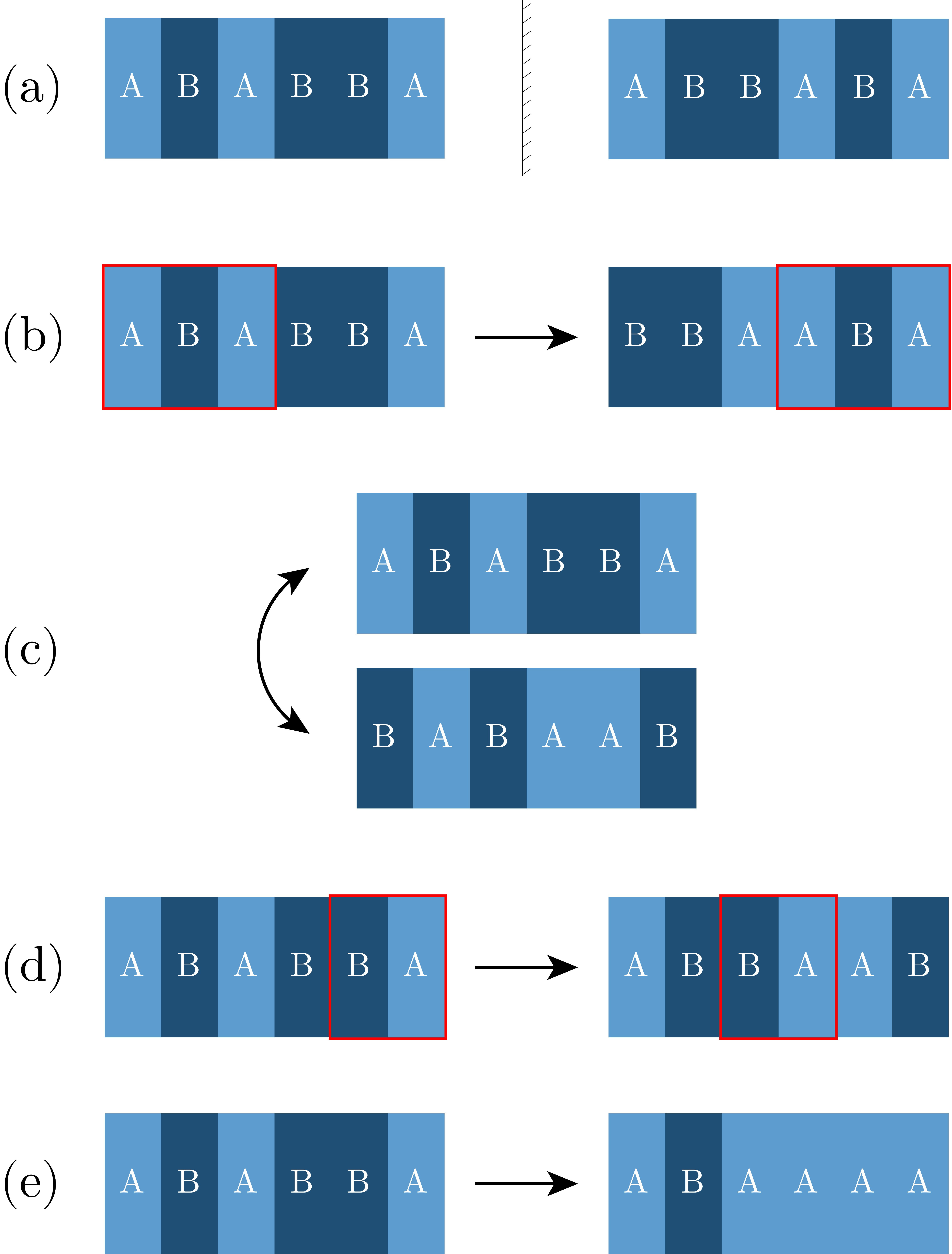}
\caption{Hidden symmetry operations that do not change $T_0$: (a) Mirror (b) Cyclic (c) Inversion (d) Double permutation (e) Pair inversion.}
\label{fig:symop}
\end{figure}

In the previous subsection, we discuss how the mirror of a sequence produces the same $T$ spectrum as the original sequence for all frequency, and so in particular, it also produces the same $T_0$ value. Mirror symmetry is schematically represented in Fig.~\ref{fig:symop}(a), and can explain how 12 of the 20 sequences (in 6 pairs) are related. This symmetry exists in both even and odd $N$. The rest of the symmetries are valid only at $f_0$ and for even $N$. 

The first of the hidden symmetries is conjectured by looking at groups of sequences like AAAABB, AAABBA, and AABBAA. These are cyclic permutations of one another, so we call this ``cyclic symmetry''. A schematic representation is seen in Fig.~\ref{fig:symop}(b). Straight away, this is a more complex form of symmetry than mirror or inversion, because it's not a single symmetry operation. Rather, it's a set of symmetries dependent on the number of layers you cycle, from 1 to $N-1$ (cycling $N$ layers would be the identity operation). The invariance of $T_0$ under cyclic symmetry can be proven directly and rather elegantly from properties of the transfer matrix, as shown in Appendix B2. 

The next symmetry is swapping all A's with B's and vice versa, which we call ``inversion symmetry'', and schematically represented in Fig.~\ref{fig:symop}(c). Inversion symmetry can explain how sequences like AABAAB and BBABBA, which seem very different at first glance, are related to each other. It turns out that this symmetry is very difficult to prove directly from operations on the transfer matrices, which is how mirror and other symmetries are proven, and we are not able to do it. Instead, the proof of inversion symmetry becomes trivial with the formula for $T_0$ given in subsection C.

Having cyclic symmetry, along with mirror and inversion, the 20 sequences that gives $T_0=1.0$ can be split into three ``cycles'', represented by AAAAAA, AAAABB, and AABAAB, such that any two sequences in the same cycle can be related by at most two of these symmetry operations. An example would be AAABBA and BBAABB, both in the cycle AAAABB, and related by an inversion and cyclic permutation. However, we still have no symmetry operation to related sequences from different cycles.

The last two symmetries essentially explain how to connect between these ``cycles''. They are rather unusual, in that there aren't analogous operations in discussing, for example, point symmetry group of molecules, which does not change numbers of A and B. The first is arbitrary permutations of double layers, that is, break up the sequence into $N/2$ two-layer segments (remember this symmetry is only for even $N$), such as (AA)(BA)(AB), and permute those segments arbitrarily, shown in Fig.~\ref{fig:symop}(d). We may call this ``double permutations''. This along with cyclic permutation allows us to get between AAAABB cycles and AABAAB as follows: we permute (AA)(BA)(AB) to (AA)(AB)(BA), then cyclical permute one place to the right to AAAABB. The proof that double permutation does not change $T_0$ is given in Appendix B3.

The last symmetry needed is inversion of a pair of layers that are the same either (AA) or (BB), which we call ``pair inversion'', shown in Fig.~\ref{fig:symop}(e). This is easy to understand, and allows us to get between AAAABB and AAAAAA, where the BB inverted to AA. AABAAB can also be turned in AAAABB by inverting the second AA in the first sequence so that it becomes AABBBB, inverting the whole sequence to BBAAAA, the cyclically permuting two places to the left. Like for inversion symmetry, the proof that pair inversion does not change $T_0$ is trivial once we define the formula for $T_0$ seen in subsection C.

These symmetry operations and their products can now relate any two sequences with even $N$ that have the same $T_0$ value. However, we have not yet investigated what those actual $T_0$ values are. In the next section, an analytical formula for $T_0$ shall be derived, for both even and odd $N$, as functions of a parameter associated with each sequence we shall call ``charge''.

\subsection{Formula for $T_0$}
Looking at all the sequences for even $N$, with different $T_0$ values, two obvious patterns immediately jump out. The two unvarying sequences (AAAA... and BBBB...) always have the highest $T_0=1.0$. On the other hand, the two alternating sequences (ABAB... and BABA...), and only those, always have the lowest $T_0$, decreasing as $N$ increases. This structure is known as a dielectric mirror or Bragg reflector, since having the lowest $T_0$ means it has the highest $R=1-T$~\cite{orfanidis2002electromagnetic}.
This is the starting point and clue that lead to the theory of ``charge'' for sequences in general. The definition of ``charge'' can be drawn by considering that the unvarying sequence can be thought of as being composed of blocks of AA or BB repeated. Similarly, the alternating sequences are blocks of AB or BA repeated. These are the two extremes, and every sequence can be thought of as being composed of some combination of these four blocks, observing that their $T_0$ values fall somewhere in between as well.

\begin{figure}[t]
\centering\includegraphics[width=8cm]{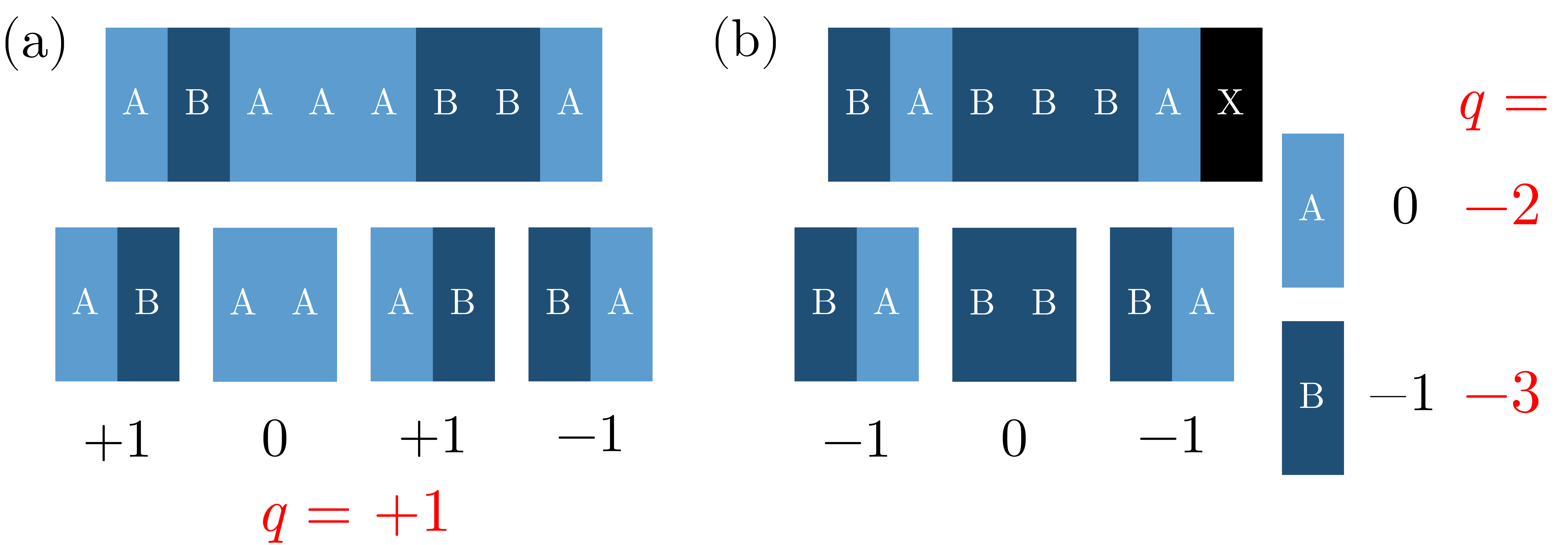}
\caption{Concept of ``charge''. (a) We assign a charge $+1,0,-1$ for AB, AA (or BB), BA sequences. Total charge $q$ is defined by the sum of charge. (b) In the case of odd number $N$, we add $0$ or $-1$ for the last $L_i$ of A and B to get $q$.}
\label{fig:charge}
\end{figure}

The basic idea is that we assign a ``charge'' to each of these blocks: AB is $+1$, BA is $-1$, and AA and BB are both $0$ as shown in Fig.~\ref{fig:charge}(a). Note that this is not inherently related to electrical charge (though we are investigating a potential link), but the trichotomy of values and, as will be seen, the behavior of charges ``cancelling'' is entirely analogous, so ``charge'' is an apt name. Given these assignments, the total ``charge'' of an even $N$ sequence is straightforwardly defined by adding together the charge of each 2-layer block, as illustrated in Fig.~\ref{fig:charge}(a). We shall denote the total ``charge'' of a sequence as $q$.

Then it can be seen that all sequences with the same $T_0$ also have the same $|q|$ as shown in Table~\ref{tab:6seq}. For example, the 20 prototypical 6-layer sequences with $T_0=1.0$ have $|q|=0$. Moreover, each $|q|$ value corresponds of only one $T_0$ value. We now see that all of the symmetries operations discussed in the previous section simply preserve $|q|$. In fact, any operation defined in subsection B on a sequence that doesn't change $|q|$ would be a symmetry operation that doesn't change $T_0$.

We know that the lowest $|q| = 0$ gives the highest $T_0=1.0$. We also know that the highest $|q|=N/2$, corresponding to the alternating sequences, gives the lowest $T_0$. Through some intuition and careful algebraic manipulation, the following formula for $T_0$ as a function of $|q|$ (and fixed $\varepsilon_{\text{A}}$, $\varepsilon_{\text{B}}$) was derived. 

\begin{equation}
T_0 = \frac{4(\varepsilon_{\text{A}}\varepsilon_{\text{B}})^{|q|}}{\left(\varepsilon_{\text{A}}^{|q|}+\varepsilon_{\text{B}}^{|q|}\right)^2}
\label{eq:formulaT}
\end{equation}

A full proof is given in Appendix C. This equation can be considered a generalization of the equation for reflectance given by Orfanidis in Chap. 6 of \cite{orfanidis2002electromagnetic}, wherein only the max $|q|$ is considered. With $|q|=0$, $T_0 = 4(1)/(1+1)^2 = 1.0$, as expected. As $|q|$ increases, the square in the denominator makes it increase faster than the numerator, meaning $T_0$ decreases.

Proof of inversion symmetry follows as a direct consequence, since Eq.~(\ref{eq:formulaT}) is symmetric with respect to $\varepsilon_{\text{A}}$ and $\varepsilon_{\text{B}}$. Pair inversion is simply replacing AA with BB and vice versa, which has no effect on $q$ and thus $T_0$, since both have 0 ``charge''. It is also clear now why there are $(N/2 + 1)$ $T_0$ values for even $N$, as conjectured. Given even $N$ layers, each sequence is composed of $N/2$ blocks, so the maximum $|q|$ is $N/2$. Every integer from 0 to $N/2$ is a possible $|q|$ value, so there are $N/2+1$ values, each corresponding to a different $T_0$ value.

We finally tackle the question of odd $N$ sequences. We found that the patterns are too difficult and non-obvious to study in terms of symmetry. However, the theory of ``charge'' offers a simpler yet more powerful tool to understand the patterns. First, we extend the definition of total ``charge'' $q$ to odd $N$. The first $N-1$ even number of layers can have ``charge'' assigned exactly as for even $N$ sequences. All that remains is one extra layer, which is either A or B. We assign a ``charge'' of 0 to A and $-1$ to B, which is added on to the ``charge'' of the first $N-1$ layers, to get the total ``charge'' $q$ as shown in Fig.~\ref{fig:charge}(b). (Note that the assignment of 0 and $-1$ was somewhat arbitrary, it could also work with 1 and 0, but the formula below would be slightly different.)

The formula for $T_0$ for odd $N$ is somewhat more tricky, but we get the following expression, where proof is given in Appendix C.

\begin{equation}
T_0 = \frac{4\varepsilon_{\text{A}}^{|q+1|}\varepsilon_{\text{B}}^{|q|}}{\left(\varepsilon_{\text{A}}^{|q+1|}+\varepsilon_{\text{B}}^{|q|}\right)^2}
\label{eq:formulaodd}
\end{equation}

We first note that it is not symmetric with respect to $\varepsilon_{\text{A}}$ and $\varepsilon_{\text{B}}$, explaining the lack of inversion symmetry that was noticed when initially looking for patterns. Second, there is no value of $q$ for which the expression reduces to 1 for any $\varepsilon_{\text{A}}$ and $\varepsilon_{\text{B}}$ except for $\varepsilon_{\text{A}}=\varepsilon_{\text{B}}=1$ (vacuum), like $|q|=0$ with even $N$, meaning perfect transmittance is not guaranteed. The fact that we get $T_0=0.99988$ very close to 1, was purely a coincidence in our choice of $\varepsilon_{\text{A}}$ and $\varepsilon_{\text{B}}$. Relatedly, $T_0$ eventually decreases as $|q|$ increases, though not monotonically as with even $N$.

Third, the function is not even in $q$, i.e. $+|q|$ and $-|q|$ give different $T_0$ values. Finally, we can explain why there are $N+1$ $T_0$ values for odd $N$, as conjectured. The first $N-1$ layers have can have ``charge'' of $0, \pm1, \dots, \pm\frac{N-1}{2}$, a total of $2\frac{N-1}{2}+1=N$ values. The final layer either doesn't change charge (if A) or decreases it by 1 (if B). For almost all of them, decreasing by 1 simply gives the charge below, no extra $q$ values, \emph{except} for the lowest charge, $-\frac{N-1}{2}$, where decreasing by 1 produces $q = -\frac{N+1}{2}$. Hence there are ($N+1$) $q$ values, each corresponding to a different $T_0$ value.

\subsection{Degeneracy}
The last unsolved question is, for a given $N$, how many sequences there are for each $T_0$ value, which we call the degeneracy at that $T_0$ ($d_N$ for even $N$ and $d'_N$ for odd $N$). The first step to understanding the pattern is noticing a connection to Pascal's triangle and the binomial coefficients. In particular, the number of sequences at $T_0=1.0$ for every even $N$ seemed to be a central binomial coefficient: 1, 2, 6, 20, 70, $\dots$. With this as the starting point, the following combinatorial formulae were inductively derived, by calculating degeneracies at each $T_0$ for increasing $N$.

\begin{equation}
d_N(|q|) =
  \begin{cases}
    \dbinom{N}{\frac{N}{2}} & \text{for } |q| = 0 \\[10pt]
    2\dbinom{N}{\frac{N}{2}+|q|} & \text{for } |q| = 1, 2, \dots, \displaystyle\frac{N}{2}\\
  \end{cases}
\label{eq:dege}
\end{equation}

\begin{equation}
d'_N(q) =\dbinom{N}{\frac{N+1}{2}+q} \text{for } q = 0, \pm1, \dots, \pm\frac{N-1}{2}, -\frac{N+1}{2}
\label{eq:dego}
\end{equation}

Armed with our understanding of ``charge'', the proof of this becomes a problem of combinatorics. Essentially, we can count the number of ways to get $q$ in $N$ layers given the number of ways to get $q-1$, $q$, and $q+1$ in $N-2$ layers, form a recurrence relation, then relate this to the binomial coefficients. A full proof is given in Appendix D.

\subsection{Bandwidth}

In this section, we investigate the how the sequence affects the bandwidth of $T$ spectra at $f_0$. Finding sequences with the sharpest peak (i.e. narrowest bandwidth) at $f_0$ is useful for application such as optical filter. However, we will not discuss about how the optical filter is realized, rather we will discuss about the behavior of the bandwidth and the pattern that gives the sharpest $T$ spectrum.

Firstly, we are only looking at sequences with the highest $T_0$, that is, with $|q| = 0$ for even $N$-layer sequences, where $T_0=1.0$. For odd $N$, there is in general no value of $q$ that gives $T_0= 1.0$. The ``charge'' $q$ that gives the highest $T_0$ varies as a function of $\varepsilon_{\text{A}}$ and $\varepsilon_{\text{B}}$. To find this $q$, we differentiate Eq.~(\ref{eq:formulaodd}) ($T_0$ for odd $N$) to find the maximum, and get
\begin{equation}
q_{\text{max}}=\text{round}\left(\frac{\log(\varepsilon_{\text{A}})}{\log(\varepsilon_{\text{B}}/\varepsilon_{\text{A}})}\right),
\end{equation}
rounding because $q$ can only take integer values. Then after setting $\varepsilon_{\text{A}}$ and $\varepsilon_{\text{B}}$, we consider only sequences with this $q_{\text{max}}$ for odd $N$. 

\begin{figure}[h]
\centering\includegraphics[width=8cm]{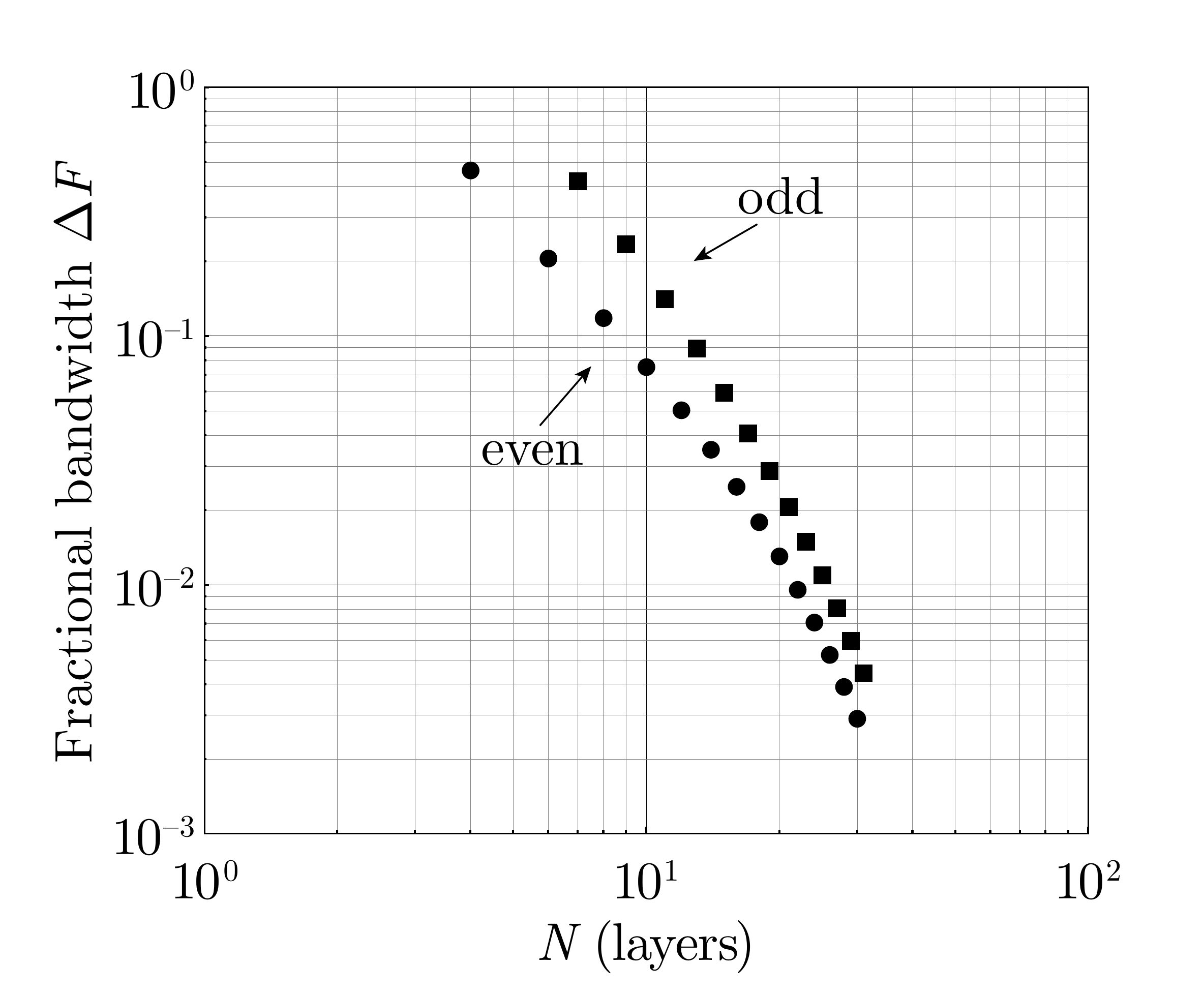}
\caption{The minimum $\Delta F$ amongst $T$ spectra that give the highest $T_0$, as a function of $N$ for even (dot) and odd (box) $N$.}
\label{fig:band}
\end{figure}

We define fractional bandwidth of a spectrum normalized to $f_0$ as $\Delta F=\Delta f/f_0$, where $\Delta f$ is the full width at half maximum (FWHM). In Fig.~\ref{fig:band} we plot the minimum $\Delta F$ for each $N$ as a function of $N$ on a log-log plot. The first thing to notice is that the minimum $\Delta F$ (the narrowest spectrum) decreases with increasing $N$. So on a very general level, to get a sharper peak in $T$ spectrum at $f_0$, we need to have more layers, as can be expected.

\begin{figure}[h]
\centering\includegraphics[width=8.5cm]{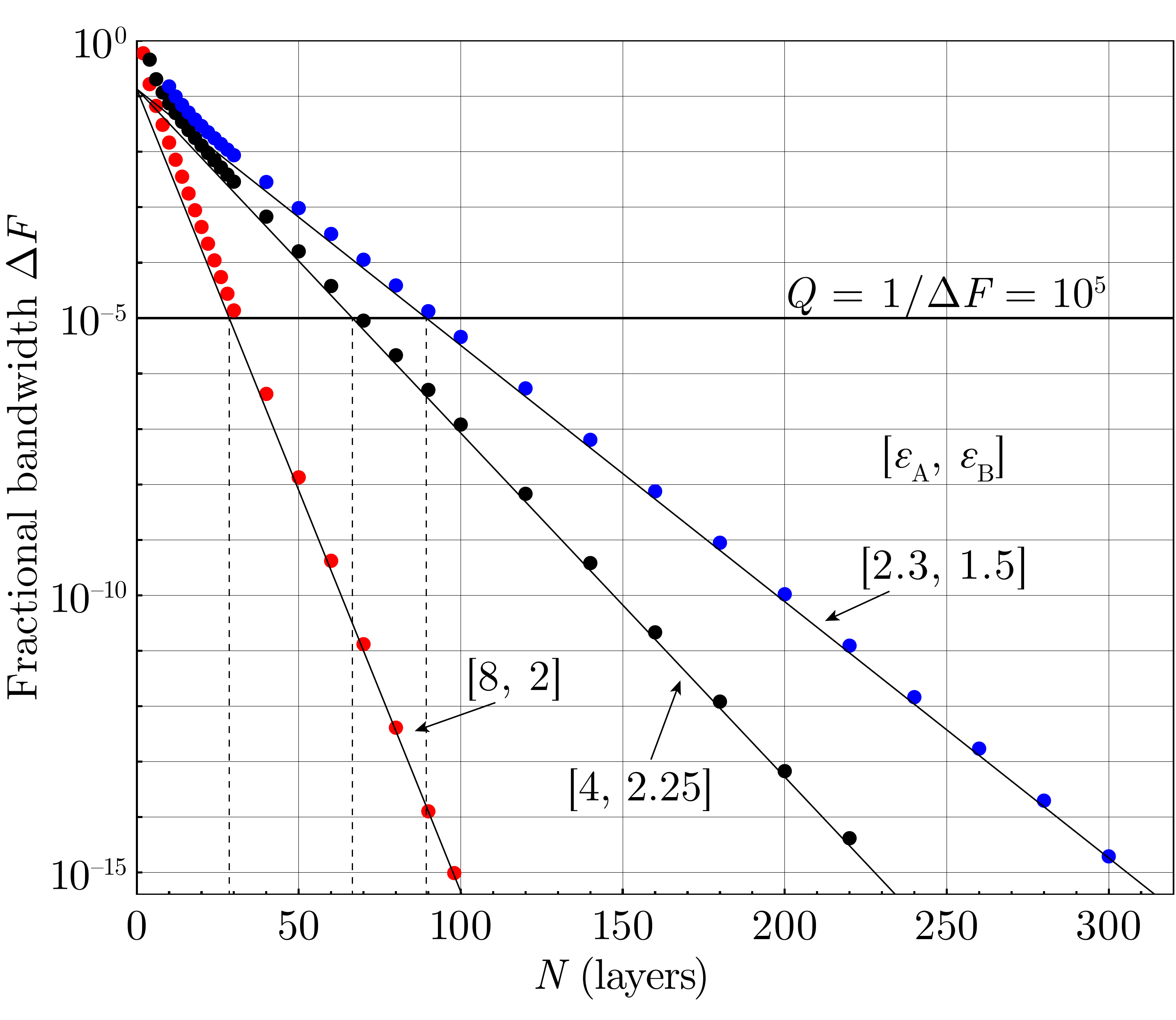}
\caption{The long-term behavior of minimum $\Delta F$ as $N$ increases. Note the change from log-log scale in Fig.~\ref{fig:band} to log-linear scale. The different colors indicate different choices of $\varepsilon_{\text{A}}$ and $\varepsilon_{\text{B}}$ shown in the figure. Quality factor $Q$, a more commonly used as a measure of bandwidth, is defined as $f_0/\Delta f = 1/\Delta F$.}
\label{fig:linband}
\end{figure}

At first, the $\Delta F$ values appear to follow roughly a straight line on the log-log plot, indicating a power law relationship. However, even in Fig.~\ref{fig:band}, the points clearly start to curve. So $\Delta F$ is calculated for larger $N$ and shown in Fig.~\ref{fig:linband} as a log-linear plot. The values of $\varepsilon_{\text{A}}$ and $\varepsilon_{\text{B}}$ were also varied to see how $\Delta F$ changes. As seen in Fig.~\ref{fig:linband}, the linear relationship between $\Delta F$ and $N$ in log-linear plot immediately jumped out, indicating exponential relationship in the linear plot. We plot only even $N$ for clarity, as odd $N$ has the same long-term linear behavior, parallel to even $N$ but shifted upwards slightly.

Although the points clearly don't follow a straight line for small $N$, they do show very regular behavior as $N$ gets larger. The exponential fit (with $\varepsilon_{\text{A}}=n_{\text{A}}^2$ and $\varepsilon_{\text{B}}=n_{\text{B}}^2$ fixed) for the asymptotic behavior was found to be the very simple equation:
\begin{equation}
\Delta F = e^{-\rho N-2},
\label{eq:df}
\end{equation}
where
\begin{equation}
\rho = \frac{n_{\text{A}}-n_{\text{B}}}{n_{\text{A}}+n_{\text{B}}}
\end{equation}
is called the (elementary) reflection coefficient~\cite{orfanidis2002electromagnetic}.

Several things may be taken from this. First, it is useful in the design of optical filters. For example, if we need a filter with a specified $Q$ factor of $10^5$ (i.e. $\Delta F = 10^{-5}$), and knowing the refractive indices $n_{\text{A}}$ and $n_{\text{B}}$ of the materials we have available and thus $\rho$, we can easily solve Eq.~(\ref{eq:df}) for an estimate of the minimum number of layers $N$ required. This is graphically represented by the dotted lines in Fig.~\ref{fig:linband}. Second, it demonstrates and moreover explains why having materials with a greater difference in refractive index is better for narrower filters, since that maximizes $\rho$, thereby minimizing $\Delta F$ for a given $N$.

It is worth pointing out that an expression for bandwidth for this type of filter is given by Macleod in \cite{macleod2001thin}. However, our expression is considerably simpler, making it much easier and quicker to solve for $N$, the only tradeoff being a worse fit for small $N$. We would like to emphasize that Eq.~(\ref{eq:df}) is an ``empirical" fit, however, its simplicity and accuracy suggests it should be possible to derive analytically with some suitable approximations to account for its asymptotic nature. Though we offer no such derivation in this paper, we would conjecture it can be derived from the expressions given in \cite{macleod2001thin}.

We also found the pattern for which sequence gives the narrowest $\Delta f$ for any given even N. It is explained in Table~\ref{tab:patt}. This pattern holds for any $\varepsilon_{\text{A}}$ and $\varepsilon_{\text{B}}$ with $\varepsilon_{\text{A}} > \varepsilon_{\text{B}}$ (otherwise simply swap A'€™s and B'€™s).

\begin{table}[h]
\begin{tabular}{c|c}
\multicolumn{2}{c}{Even $N$}\\
\hline\hline
\multicolumn{1}{c}{\hspace*{4cm}} & \multicolumn{1}{c}{\hspace*{4cm}} \\[-10.5pt]
If $\displaystyle\frac{N}{2}$ is even: & If $\displaystyle\frac{N}{2}$ is odd: \rule{0pt}{14pt}\rule[-10pt]{0pt}{0pt}\\
$\underbrace{\text{AB...AB}}_{\frac{N}{4} \times \text{AB}}\underbrace{BA...BA}_{\frac{N}{4} \times \text{BA}}$ & $\underbrace{\text{AB...AB}}_{\frac{N-2}{4} \times \text{AB}}\text{AA}\underbrace{\text{BA...BA}}_{\frac{N-2}{4} \times \text{BA}}$ \rule[-20pt]{0pt}{0pt}\\
\hline
E.g. $N$ = 8: $\displaystyle\frac{N}{2} = 4$ & E.g. $N$ = 14: $\displaystyle\frac{N}{2} = 7$ \rule{0pt}{14pt}\rule[-10pt]{0pt}{0pt}\\
$\underbrace{\text{ABAB}}_{2 \times \text{AB}}\underbrace{BABA}_{2 \times \text{BA}}$ & $\underbrace{\text{ABABAB}}_{3 \times \text{AB}}\text{AA}\underbrace{\text{BABABA}}_{3 \times \text{BA}}$ \rule[-16pt]{0pt}{0pt}\\
\hline\hline
\end{tabular}
\caption{Pattern for the sequence that gives the narrowest $\Delta F$ for even $N$.}
\label{tab:patt}
\end{table}

It may be of interest to note that the second narrowest sequence for any given even $N$ follows a very simple pattern too. Simply replace the middle two layers with AA if it's BB, and vice versa. For example, for $N = 8$, the narrowest spectrum is given by the sequence ABA\textbf{BB}ABA, the second narrowest is given by ABA\textbf{AA}ABA. These in fact exactly correspond to the high-index and low-index cavity all-dielectric filters described by Macleod in \cite{macleod2001thin}, and we have now conclusively shown, by calculating all $2^N$ sequences, that they are the ``best" possible filters (in terms of bandwidth) for a given $N$.

A similar albeit more complicated pattern was found for odd $N$. However, because the $q$ that gives the highest $T_0$ varies as a function of $\varepsilon_{\text{A}}$ and $\varepsilon_{\text{B}}$, so too does this pattern. Thus, we feel it is not worth describing here the rule for odd $N$, since it only works for some particular values of $\varepsilon_{\text{A}}$ and $\varepsilon_{\text{B}}$, along with the fact that the narrowest bandwidth for any odd $N$ is larger than that for the even $N-1$.

\section{Conclusion}

In conclusion, we have found that, somewhat unexpectedly, the transmittance $T$ of $N$-layer dielectric stacks are highly degenerate and discrete at the central frequency $f_0$. We have found all hidden symmetry operations to sufficiently explain how all even $N$ sequences with the same $T_0$ are related. Furthermore, $T_0$ depends only on the ``charge'' $q$ of a sequence, with formulae for $T_0$, for both even and odd $N$, derived as functions of $q$. This is a simpler, more elegant way to explain why different sequences have the same $T_0$ value. The degeneracy at each $T_0$ is explained by combinatorics, again with formulae derived as functions of $q$.

There is a lot of potential for future work, in various directions stemming from this initial discovery and investigation. A well-established mathematical tool to analyze and understand symmetries is group theory. In fact, we have already started in this endeavor, trying to form a group of symmetry operations both for $N=2$ and for $N=4$, then analyzing the structure using representation theory to extract the degeneracies. However, we run into issues such as not being able to include some of the more exotic operations in the group, and the predicted degeneracies of irreducible representation do not match the degeneracies that we calculated. These problem might be related to the fact that we discuss transmittance but not transmission coefficient or any eigenvalue of linear operators that commute with symmetry operations. Alternatively, the symmetry operations could potentially have the structure of a groupoid, a generalization of a group.

Recalling that PCs may be used for optical filters, we want sequences with both high $T$ \emph{and} a sharp peak at $f_0$. Now that we understand how to find $T_0$ just by looking at the sequence, we only have to consider a much smaller subset of sequences, those with low $q$ and high $T_0$. The next big step is to continue our preliminary investigation into how bandwidth depends on the sequence, e.g. we would want a sharp peak for a filter. If we can fully understand how that changes under the symmetry operations as well, we could imagine creating an algorithm to find the optimal sequence for any kind of $T$ spectrum desired for a given $N$, or designing sequences satisfying some given requirements (e.g. $Q$ factor $>$ some value).

An interesting and potentially fruitful area to investigate is whether there is any physical meaning to this artificial value associated with a sequence we call ``charge''. Again bringing it back to physical applications, PCs can also have $E$ field enhancement, which is useful in the enhanced Raman spectroscopy to get a stronger signal. Preliminary investigations suggest that there may be a relationship between ``charge'' or ``cumulative charge'' in the sequence, and the $E$ field within the PC. For example, a sequence like ABABAB...BABABA has overall $|q|=0$ so $T_0=1.0$. But right at the middle of the sequence, it has very high ``cumulative charge'', and correspondingly, a very high $E$ field at the middle point. Further investigation and understanding could allow us to design PC sequences with $E$ field enhancement at any position we desire.

Before finishing the story, we would like to point out similarities of the present story to a general physics in which odd and even number of particles give a different symmetry (or statistics). Although it is beyond our ability, it is our pleasure if the reader has an interest in such hidden symmetries for applying to general physics.

\begin{acknowledgments}
H. L. thanks J. Kono, C. J. Stanton, S. Phillips, K. Packard, K. Ogawa, and U. Endo for 
making the Nakatani RIES program possible.
M.S.U. is supported by the MEXT scholarship.
R.S. acknowledges JSPS KAKENHI Grant Numbers JP 25107005 and JP 25286005.
\end{acknowledgments}


\appendix
\section{}
\textbf{Appendix omitted in this version (waiting until publication).}

\bibliography{stda2000}

\end{document}